\documentclass[twocolumn,aps]{revtex4}
\usepackage{amsmath}
\usepackage{hyperref}
\hypersetup{colorlinks=true,linkcolor=blue,citecolor=red}
\begin{document}
\title{Appropriate Inner Product for PT-Symmetric Hamiltonians}
\author{Philip D. Mannheim\\
Department of Physics, University of Connecticut, Storrs, CT 06269, USA\\
email: philip.mannheim@uconn.edu}
\date{December 10, 2017}

\begin{abstract}
A Hamiltonian $H$ that is not Hermitian can still have a real and complete energy eigenspectrum  if it instead is $PT$ symmetric. For such Hamiltonians three possible inner products have been considered in the literature, the $V$ norm, the $PT$ norm, and the $C$ norm. Here $V$ is the operator that implements $VHV^{-1}=H^{\dagger}$, the $PT$ norm is the overlap of a state with its $PT$ conjugate, and $C$ is a discrete linear operator that always exists for any Hamiltonian that can be diagonalized. Here we show that it is the $V$ norm that is the most fundamental as it is always chosen by the theory itself. In addition we show that the $V$ norm is always equal to the $PT$ norm if one defines the $PT$ conjugate of a state to contain its intrinsic $PT$ phase. We discuss the conditions under which the $V$ norm coincides with the $C$ operator norm, and show that in general one should not use the linear $C$ operator but for the purposes that it is used one can instead use the antilinear $PT$ operator itself.
\end{abstract}
\maketitle

\section{Implications of Antilinear Symmetry}
In an analysis of the eigenspectrum of the non-Hermitian Hamiltonian $H=p^2+ix^3$ Bender and collaborators  \cite{Bender1998,Bender1999}  unexpectedly found that the eigenvalues of $H=p^2+ix^3$ are all real \cite{footnote1A}. The reality of the eigenvalues was traced to the fact that while not Hermitian the $p^2+ix^3$ Hamiltonian had an antilinear $PT$ symmetry, where $P$ denotes parity and $T$ denotes time reversal. (Under $PT$: $p\rightarrow p$, $x\rightarrow -x$, $i \rightarrow -i$, so that $p^2+ix^3\rightarrow p^2+ix^3$.) While sufficient to secure the reality of eigenvalues Hermiticity  is thus seen as not being necessary for such reality. In fact  \cite{Bender2002,Bender2010} it was antilinearity that emerged as the necessary condition for the reality of eigenvalues, so that without an antilinear symmetry energy eigenvalues could not all be real. Following on from the work of \cite{Bender1998,Bender1999} there has been much interest in the literature (see e.g. the reviews of  \cite{Bender2007,Special2012,Theme2013} and the studies of \cite{Mostafazadeh2002,Solombrino2002}) in such antilinear symmetry, as it was realized that one can actually replace the familiar postulate of Hermiticity of a Hamiltonian by the more general requirement of antilinear symmetry (antilinearity) without needing to either generalize or modify the basic structure of quantum mechanics in any way. Moreover, antilinearity was actually shown \cite{Mannheim2015b} to be the most general requirement that one could impose on a quantum theory for which it would continue to be viable, since in addition to antilinear symmetry being necessary for the reality of eigenvalues,  antilinear symmetry is also necessary for the  existence of  a sensible Hilbert space description of quantum mechanics wherein one is able to define an inner product that is time independent, positive definite, and finite. There is however no need for the inner product to be composed of a ket and its Hermitian conjugate bra or for the Hamiltonian to be Hermitian in order to achieve this, and through study of various $PT$-symmetric examples some possible inner products have been identified in the literature that could achieve this objective. It is the purpose of this paper to elucidate the connections between these various inner products, and identify one of them (the so-called $V$ norm discussed in \cite{Mostafazadeh2002,Solombrino2002}) to be the most general in that it must be possessed by any theory with antilinear symmetry whose Hamiltonian is diagonalizable.

To understand the implications of  antilinear symmetry for a Hamiltonian $H$, it is instructive to consider the eigenvector equation obeyed by its eigenvectors:
\begin{eqnarray}
i\frac{\partial}{\partial t}|\psi(t)\rangle=H|\psi(t)\rangle=E|\psi(t)\rangle.
\label{N1}
\end{eqnarray}
On replacing the parameter $t$ by $-t$ and then multiplying by some  general antilinear operator $A$ we obtain
\begin{eqnarray}
i\frac{\partial}{\partial t}A|\psi(-t)\rangle=AHA^{-1}A|\psi(-t)\rangle=E^*A|\psi(-t)\rangle.
\label{N2}
\end{eqnarray}
From (\ref{N2}) we see that if $H$ has an antilinear symmetry so that $AHA^{-1}=H$, then, as first noted by Wigner in his study of time reversal invariance,  energies can either be real and have eigenfunctions that obey $A|\psi(-t)\rangle=|\psi(t)\rangle$, or can appear in complex conjugate pairs that have conjugate eigenfunctions ($|\psi(t)\rangle \sim \exp(-iEt)$ and $A|\psi(-t)\rangle\sim \exp(-iE^*t)$). Antilinearity thus admits of two possibilities, energies all real, or some or all of the energies appearing in complex conjugate pairs. 

It is also possible to establish a converse theorem. Thus suppose we are given that the energy eigenvalues are real or appear in complex  conjugate pairs. In such a case not only would $E$ be an eigenvalue but $E^*$ would be too. Hence, we can set $HA|\psi(-t)\rangle=E^*A|\psi(-t)\rangle$ in (\ref{N2}), and obtain
\begin{eqnarray}
(AHA^{-1}-H)A|\psi(-t)\rangle=0.
\label{N3}
\end{eqnarray}
Then if the eigenstates of $H$ are complete, (\ref{N3}) must hold for every eigenstate, to yield $AHA^{-1}=H$ as an operator identity, with $H$ thus having an antilinear symmetry. (We can use the standard argument based on completeness for linear operators here since while $A$ itself might be antilinear the operator $AHA^{-1}$ is linear.) Antilinearity is thus both necessary and sufficient for energy eigenvalues  to be real or appear in complex conjugate pairs, and thus without antilinearity it is not possible for all eigenvalues to be real.

Now we note that there is no analog statement for Hermiticity, since while Hermiticity implies the reality of eigenvalues, there is no converse requirement that the reality of eigenvalues implies Hermiticity. It is in this regard that antilinear symmetry is more general than Hermiticity while encompassing it as a special case.

To  illustrate the basic features of antilinear symmetry it is convenient to consider  a very simple model, viz. the matrix 
\begin{eqnarray}
M(\alpha,\beta)=\alpha\sigma_1+i\beta\sigma_2=\begin{pmatrix}0&\alpha+\beta\\ \alpha-\beta&0\\
\end{pmatrix},
\label{N4}
\end{eqnarray}
where the parameters $\alpha$ and $\beta$ are real and positive. (We shall have occasion to return to this model below.) The matrix $M(\alpha,\beta)$ does not obey the transposition plus complex conjugation Hermiticity condition $M_{ij}=M^*_{ji}$. However, if we set $P=\sigma_1$ and  $T=Ki\sigma_1$, where $K$ denotes complex conjugation, we obtain $PTM(\alpha,\beta)T^{-1}P^{-1}=M(\alpha,\beta)$, with $M(\alpha,\beta)$ thus being  $PT$ symmetric for any values of  the real parameters $\alpha$ and $\beta$. With the eigenvalues of $M(\alpha,\beta)$ being given by $E_{\pm}= \pm (\alpha^2-\beta^2)^{1/2}$, just as required we see that both of these eigenvalues are real if $\alpha \geq \beta$, and form a complex conjugate pair if $\alpha<\beta$. And while the energy eigenvalues would be real and degenerate (both eigenvalues being equal to zero) at the crossover point where $\alpha=\beta$, at this point the matrix becomes of non-diagonalizable Jordan-block form (see e.g. the analog discussion in \cite{Mannheim2013,Mannheim2015b}). Neither of the $\alpha = \beta$ or $\alpha < \beta$  possibilities are achievable with Hermitian Hamiltonians, while for $\alpha > \beta$ the matrix $M(\alpha > \beta)$ is an example of a non-Hermitian matrix that has real eigenvalues \cite{footnote2A}.

While our analysis here will focus specifically on $PT$ symmetry itself, as far as non-relativistic quantum mechanics is concerned our analysis could be applied to any antilinear symmetry. However, once one considers the implications of relativity, the requirement of the time independence of inner products coupled with the imposition of complex Lorentz invariance leads us uniquely to one specific antilinear symmetry, namely $C_CPT$ \cite{Mannheim2015b,Mannheim2016b}, where $C_C$ denotes the discrete charge conjugation operator that squares to one. (The operator $C_C$ is different from the operator $C$ \cite{Bender2007} that will be discussed in detail below, an operator that both commutes with $H$ and squares to one, and always exists since in the basis in which $H$ is diagonal one can always find other diagonal operators with arbitrary eigenvalues that commute with $H$.) The work of \cite{Mannheim2015b,Mannheim2016b} thus generalizes the $CPT$ theorem to the non-Hermitian case. If however, we work below the threshold for pair creation charge conjugation will play no role with $C_C$ then commuting with $H$, and thus in the following we shall restrict to just $PT$ symmetry itself \cite{footnote3A}. In order to discuss and compare some candidate inner products for $PT$-symmetric theories we shall first discuss those realizations of $PT$ symmetry in which all energy eigenvalues are real, and turn now to the $V$ norm.

\section{The $V$ norm}

For our discussion of the $V$ norm and in order to be able to compare and contrast the various inner products that have been discussed in the PT literature, it suffices to restrict the discussion to  Hamiltonians that have an antilinear symmetry, that do not obey $H=H^{\dagger}$, but have an energy eigenspectrum that is complete  \cite{footnote4A}. We thus restrict to Hamiltonians that act on the same kind of Hilbert spaces with complete and normalizable bases of eigenvectors as Hermitian operators do, with the Hamiltonians of interest to us here being diagonalizable \cite{footnote5A}. We first discuss the case of diagonalizable Hamiltonians with eigenvalues that are all real, and below we consider the other realization of antilinear symmetry, namely eigenvalues that come in complex conjugate pairs. 

For the real energy case the Hamiltonian can be brought to a Hermitian form by a similarity transform $SHS^{-1}=H^{\prime}$ in which $H^{\prime}$ obeys $H^{\prime}=H^{\prime \dagger}$. The eigenstates of $H$ and $H^{\prime}$ obey
\begin{eqnarray}
i\partial_t|R_n\rangle&=&H|R_n\rangle=E_n|R_n\rangle,
\nonumber\\
-i\partial_t\langle R_n|&=&\langle R_n|H^{\dagger}=\langle R_n|E_n,
\label{N5}
\end{eqnarray}
\begin{eqnarray}
 i\partial_t|R_n^{\prime}\rangle&=&H^{\prime}|R_n^{\prime}\rangle=E_n|R_n^{\prime}\rangle,\nonumber\\
 -i\partial_t\langle R_n^{\prime}|&=&\langle R_n^{\prime}|H^{\prime\dagger}=\langle R_n^{\prime}|H^{\prime}=\langle R_n^{\prime}|E_n,
 \label{N6}
\end{eqnarray}
and are related by
\begin{eqnarray}
 |R_n^{\prime}\rangle=S|R_n\rangle,~~~\langle R_n^{\prime}|=\langle R_n|S^{\dagger}.
\label{N7}
\end{eqnarray}
On normalizing the eigenstates of $H^{\prime}$ to unity,  we obtain
\begin{eqnarray}
\langle R_n^{\prime} |R_m^{\prime}\rangle=\langle R_n|S^{\dagger}S|R_m\rangle=\delta_{m,n}.
\label{N8}
\end{eqnarray}

The $\langle R_n^{\prime} |R_n^{\prime}\rangle$ norm is a conventional time-independent,  positive-definite Hermitian theory norm, and if $H$ can be brought to a Hermitian form by a similarity transform then the $\langle R_n|S^{\dagger}S|R_m\rangle$ norm is the norm to use for $H$, and it is both time independent and positive definite too. 

Since via the similarity transformation $S$ we can bring $H$ to a form  $SHS^{-1}=H^{\prime}$ in which $H^{\prime}$ obeys $H^{\prime}=H^{\prime \dagger}$, we thus obtain
\begin{eqnarray}
SHS^{-1}=S^{\dagger -1}H^{\dagger}S^{\dagger},
\label{N9}
\end{eqnarray}
and on introducing $V=S^{\dagger}S$ obtain
\begin{eqnarray}
VHV^{-1}=H^{\dagger}.
\label{N10}
\end{eqnarray}
The $V=S^{\dagger}S$ operator thus serves as what is known as an intertwining operator (it intertwines a Hamiltonian with its Hermitian conjugate) of the type that had been discussed in \cite{Mostafazadeh2002,Solombrino2002} (see also \cite{Mannheim2013,Mannheim2015b}). And as we see, $V$ does not just obey $V^{\dagger}=V$,  in addition $V$ is also a positive operator (i.e. all of its eigenvalues are positive) of the type discussed in \cite{Mostafazadeh2002}.

The transformation on $V$ in going from the $H$ system to the $H^{\prime}$ system is not a standard similarity transformation, even though one uses a standard similarity transformation to transform $H$ to $H^{\prime}$. Rather, one has to set
\begin{eqnarray}
V^{\prime}=S^{-1 \dagger }VS^{-1},
\label{N11}
\end{eqnarray}
since
\begin{eqnarray}
V^{\prime}H^{\prime}V^{\prime -1}&=&S^{-1\dagger }VS^{-1}SHS^{-1}SV^{-1}S^{\dagger}
\nonumber\\
&=&
S^{-1 \dagger }H^{\dagger}S^{\dagger}=H^{\prime \dagger}.
\label{N12}
\end{eqnarray}
Thus even after transforming in this specific way, $V^{\prime}$ still transforms a Hamiltonian into its Hermitian conjugate. With this transformation, and unlike with a standard similarity transformation, one can thereby transform $V$ into a $V^{\prime}$ that can be equal to one, as such a $V^{\prime}$ would generate $V^{\prime}H^{\prime}V^{\prime -1}=H^{\prime \dagger}=H^{\prime}$, and this is the case when $V=S^{\dagger}S$.

On normalizing the eigenstates of $H^{\prime}$ to unity, and with $V$ now taken to be $S^{\dagger}S$  we obtain
\begin{eqnarray}
\langle R_n^{\prime} |R_m^{\prime}\rangle=\langle R_n|S^{\dagger}S|R_m\rangle=\langle R_n|V|R_m\rangle=\delta_{m,n},
\label{N13}
\end{eqnarray}
and thus establish that the $V$ norm is not just time independent (since $\langle R_n^{\prime}|R_m^{\prime}\rangle$ is), but also establish that the $V$ norm is automatically both positive and (ortho) normalized to one. For the unprimed system then the $V$ norm is the one we need, since it is automatically both time independent and positive definite. Thus for any $H$ that is similarity equivalent to a Hermitian $H^{\prime}$  the $V$ inner product will always exist, and it is thus the most general one that one could use. 

Since the $V$ norm is time independent we obtain:
\begin{eqnarray}
i\partial_t\langle R_n|V|R_m\rangle=\langle R_n|VH-H^{\dagger}V|R_m\rangle=0.
\label{N14}
\end{eqnarray}
Since the $|R_n\rangle$ states are complete ($H^{\prime}$ being assumed to be Hermitian), we can set
\begin{eqnarray}
VH-H^{\dagger}V=0
\label{N15}
\end{eqnarray}
as an operator identity, and with $V$ necessarily being invertible (since we had initially assumed that $S$ was), we come right back to $VHV^{-1}=H^{\dagger}$ \cite{footnote6A}. With $H$ and $H^{\dagger}$ thus being isospectrally related (which all on its own entails that the energy eigenvalues of $H$ are real or in complex conjugate pairs), it follows from our discussion above regarding the relation between antilinearity and the structure of the energy eigenspectrum that $H$ has an antilinear symmetry. We thus establish that for a Hamiltonian whose energy eigenspectrum is real and complete there must exist an intertwining operator for it that effects $VHV^{-1}=H^{\dagger}$, and in consequence the Hamiltonian must possess an antilinear symmetry. Thus we can necessarily construct a positive-definite norm $\langle R_n|V|R_m\rangle$, and more importantly can do so without needing to specify what the antilinear symmetry that $H$ has to possess might even be at all. Thus even if a Hamiltonian has an antilinear symmetry other than $PT$ one must still use the $V$ norm. By the same token we note that since the eigenvalues of a Hamiltonian with  an antilinear symmetry are either real or in complex pairs, for any Hamiltonian with an antilinear symmetry  $H$ and $H^{\dagger}$ must still be isospectrally related by a $V$ that generates $VHV^{-1}=H^{\dagger}$. Thus even when energies appear in complex conjugate pairs, we can still use the $V$ norm, a point we shall return to below.

\section{The $PT$ conjugate norm}

We can write the $\langle R_n^{\prime} |R_n^{\prime}\rangle$ norm as $\langle R_n^{\prime} |R_n^{\prime}\rangle=(|R_n^{\prime}\rangle)^{\dagger}|R_n^{\prime}\rangle$ where  the dagger denotes Hermitian adjoint. Inserting a complete set of position eigenstates then gives
\begin{eqnarray}
&&\langle R_n^{\prime} |R_n^{\prime}\rangle=(|R_n^{\prime}\rangle)^{\dagger}|R_n^{\prime}\rangle=\int dx (|R_n^{\prime}\rangle)^{\dagger}|x\rangle\langle x| R_n^{\prime}\rangle
\nonumber\\
&&=\int dx (\psi^{\prime}_n(x))^{\dagger}\psi^{\prime}_n(x)=\int dx \psi_n^{\prime *}(x)\psi^{\prime}_n(x)=1.~~~~
\label{N16}
\end{eqnarray}
So now let us do the exactly the same thing for $PT$. We had noted above that when all energies are real the eigenstates of $H^{\prime}$ are also eigenstates of $P^{\prime}T^{\prime}$, where $SPS^{-1}=P^{\prime}$, $STS^{-1}=T^{\prime}$. We can thus set  
\begin{eqnarray}
P^{\prime}T^{\prime}|R_n^{\prime}\rangle =\eta_n|R_n^{\prime}\rangle,
\label{N17}
\end{eqnarray}
where $\eta_n$ is an appropriate phase. 

Since $(PT)^2=I$ (or equivalently $(P^{\prime}T^{\prime})^2=I$), in general we can  take the $PT$ eigenvalue of a real energy eigenstate $|n\rangle$ to be $e^{i\alpha}$ where $\alpha$ is real, since $(PT)^2|n\rangle=PTe^{i\alpha}|n\rangle=e^{-i\alpha}PT|n\rangle=e^{-i\alpha}e^{i\alpha}|n\rangle=|n\rangle$. If we now define a new state $|\hat{n}\rangle=e^{i\alpha/2+i\beta}|n\rangle$ where $\beta$ is real, then $PT|\hat{n}\rangle=e^{-i\alpha/2-i\beta}PT|n\rangle=e^{-i\alpha/2-i\beta}e^{i\alpha}|n\rangle=e^{-i\alpha/2-i\beta}e^{i\alpha}e^{-i\alpha/2-i\beta}|\hat{n} \rangle=e^{-2i\beta}|\hat{n}\rangle$. Since $\beta$ is arbitrary we can always  choose the phases of the states so that their $PT$ eigenvalues are real, and since $(PT)^2=I$ we can take them to be equal to either plus one ($\beta=0$) or minus one ($\beta=\pi/2$). Finally, since $P^{\prime}$ and $T^{\prime}$  obey the same conditions as $P$ and $T$ (viz. $P^2=I$, $T^2=I$, $[P,T]=0$),  the $\eta_n$ phases can always be set equal to plus or minus one in either basis and are preserved under a similarity transformation.

Having now fixed the $\eta_n$ phases, we next need to identify $\langle R_n^{\prime} |R_n^{\prime}\rangle$ with a $PT$ conjugate-based norm. However, since $\langle R_n^{\prime} |R_n^{\prime}\rangle$ is positive definite, we will need a definition of a $PT$ conjugate that will lead to a  $PT$ conjugate-based norm that is itself positive definite. As we now show, this will require including the intrinsic $PT$ phase $\eta_n$ in the definition of the $PT$ conjugate. We thus identify $(|R^{\prime})^{PT}|x \rangle=(\psi^{\prime}_n(x))^{PT}$ not as its parity and complex conjugate $\psi_n^{\prime *}(-x)$ (the $-x$ factor is because of the presence of the $P$ operator), but as 
\begin{eqnarray}
(\psi^{\prime}_n(x))^{PT}=\eta_n^{-1}\psi^{\prime *}(-x),
\label{N18}
\end{eqnarray}
so that the $PT$ conjugate depends on the intrinsic $PT$ parity of the state and is different for states with differing intrinsic $PT$ parity. With this definition of the $PT$ conjugate we obtain 
\begin{eqnarray}
&&\langle R_n^{\prime} |R_n^{\prime}\rangle= (|R_n^{\prime}\rangle)^{PT} |R_n^{\prime}\rangle
\nonumber\\
&&=\int dx (|R_n^{\prime}\rangle)^{PT}|x\rangle\langle x| R_n^{\prime}\rangle
=\int dx (\psi^{\prime}_n(x))^{PT}\psi^{\prime}_n(x)
\nonumber\\
&&=\eta_n^{-1}\int dx \psi_n^{\prime *}(-x)\psi^{\prime}_n(x)=1,
\label{N19}
\end{eqnarray}
and thus obtain
\begin{eqnarray}
\int dx \psi_n^{\prime *}(-x)\psi^{\prime}_n(x)=\eta_n.
\label{N20}
\end{eqnarray}

Now $PT$ studies (typically in some equivalent basis such as the unprimed one) have shown that  integrals of the generic form $\int dx \psi_n^{\prime *}(-x)\psi^{\prime }_n(x)$ (in either the primed or unprimed bases) are not positive definite. The integral $\int dx \psi_n^{\prime *}(-x)\psi_n^{\prime }(x)$ is however real since 
\begin{eqnarray}
\left[\int dx\psi_n^{\prime *}(-x)\psi_n^{\prime }(x)\right]^*&=&\int dx\psi_n^{\prime *}(x)\psi_n^{\prime }(-x)
\nonumber\\
&=&\int dx\psi_n^{\prime *}(-x)\psi_n^{\prime }(x),
\label{N21}
\end{eqnarray}
and is automatically normalized to plus or minus one since $\int dx \psi_n^{\prime *}(-x)\psi_n^{\prime }(x)=\eta_n$. Thus with $\eta_n=\pm 1$,  one can always choose the signs of the $\eta_n$ so that $\eta_n^{-1}\int dx \psi_n^{\prime *}(-x)\psi_n^{\prime }(x)$ is positive definite.  Thus just as $PT$ acts on a ket to produce an intrinsic $PT$ phase as per
\begin{eqnarray}
|R_n^{\prime}\rangle=\eta_n^{-1}P^{\prime}T^{\prime}|R_n^{\prime}\rangle,
\label{N22}
\end{eqnarray}
the $PT$ conjugate of the ket should contain the same intrinsic phase as per
\begin{eqnarray}
(|R_n^{\prime}\rangle)^{PT}=\eta_n^{-1}\langle R^{\prime}_n|P.
\label{N23}
\end{eqnarray}
And when one includes this phase, $ (|R_n^{\prime}\rangle)^{PT} |R_n^{\prime}\rangle$
is just as positive definite as the $\langle R_n^{\prime} |R_n^{\prime}\rangle$ matrix element to which it is to be equal to. Finally, we note that previously in the literature \cite{Bender2007} it was $\psi_n^{\prime *}(-x)$ without the $\eta_n^{-1}$ factor that was taken to be the $PT$ conjugate. With this choice one then had to introduce some other discrete operator whose eigenvalues were plus and minus one, an operator, called $C$ in the $PT$ literature, that would commute with the Hamiltonian and generate the needed additional phase in order to yield a positive definite inner product. We shall return to a discussion of the $C$ operator below, while noting now that with the $\eta_n$-dependent definition of the $PT$ conjugate, even while it is useful the $C$ operator is not in fact needed.

\section{The $PV$ Operator}

Now the discussion given above is far as we can go on general grounds. However, we can go further if the parity operator effects $P^{-1}HP=H^{\dagger}$. This cannot in general be the case of course since it is possible for a Hamiltonian to be $P$ invariant, with $PT$ symmetry then reducing to $T$ symmetry, a symmetry that is still antilinear, and that still can apply to non-Hermitian Hamiltonians.  Many examples of non-Hermitian Hamiltonians that obey $P^{-1}HP=H^{\dagger}$ have been found in the literature, with an $H=p^2+ix^3$ with Hermitian and parity odd $x$ and $p$ being perhaps the prime example. To be more general we note first that in the canonical commutator $[x,p]=i$, if $p$ acts to the right  (on a ket) it can be represented as $p=-i\partial_x$, while  if $p$ acts to the left (on a bra)  it can be represented as $p=+i\partial_x$, i.e. it can be represented by the discrete parity transform of $-i\partial_x$, i.e. as $PpP^{-1}$ and thus as $P^{-1}pP$ since $P^2=I$. Now if there is just one coordinate then terms such as $x^2$, $xp$  and $p^2$  would transform into themselves under parity ($x$ and $p$ must have the same parity since $[x,p]=i$), while terms such as $ix^3$ and $ap$ where $a$ is independent of $x$ or $p$ would transform into minus themselves. For Hamiltonians that involve this latter case, the bra evolves with $P^{-1}HP$, and on recalling that a bra $\langle R_n|$ evolves with $H^{\dagger}$, we can set
\begin{eqnarray}
P^{-1}HP=H^{\dagger}.
\label{N24}
\end{eqnarray}

Thus just like $V$, $P^{-1}$ also transforms $H$ into $H^{\dagger}$. However, $P$ and $V$ are different. $P$ acts on individual operators independent of how they make up the total Hamiltonian while $V$ depends on the particular Hamiltonian. (For $H=i\lambda x^3$ for instance $P$ is independent of $\lambda$ but $V$ is not.) Also $P$ is to square to one while $V=S^{\dagger}S$ does not \cite{footnote7A}.

Given $P^{-1}HP=H^{\dagger}$, and thus $PH^{\dagger}P^{-1}=H$, we can set
\begin{eqnarray}
PVHV^{-1}P^{-1}=PH^{\dagger}P^{-1}=H,
\label{N25}
\end{eqnarray}
and thus establish that $H$ commutes with the operator $PV$. Thus when $P$ effects $P^{-1}HP=H^{\dagger}$, $H$ with its antilinear symmetry then has a linear symmetry also, and we note that the operator $PV$ is Hamiltonian dependent since $V$ is related to the operator $S$ that brings $H$ to a Hermitian form. 

We thus see that for any $PT$ invariant $H$ that obeys $P^{-1}HP=H^{\dagger}$, we can always find a linear operator $PV$ that commutes with $H$. However, this $PV$ operator can not in general be identified with the discrete linear $C$ operator that also commutes with $H$ since unlike $C$, $PV$ is not required to obey $(PV)^2=I$ and have eigenvalues equal to plus or minus one. (To obtain $PVPV=I$ one would need $PVP=V^{-1}$, which would not necessarily hold in general.) However, since $PV$ commutes with $H$ the eigenstates of $H$ are also eigenstates of $PV$, and we can take their eigenvalues to be $\alpha_n$, with these $\alpha_n$ not only not needing to be plus or minus one, they do not (initially at least) even need to be real, since even if  $P$ and $V$ are both Hermitian, they do not in general commute, and so $PV$ is not necessarily Hermitian. (In the primed basis $\sum |R_n^{\prime}\rangle \alpha_n\langle R_n^{\prime}|$ commutes with $H^{\prime}$ for any choice of the $\alpha_n$, and this is of course the reason why $C^{\prime}=\sum |R_n^{\prime}\rangle c_n\langle R_n^{\prime}|$ with $c_n=\pm 1$ commutes with $H^{\prime}$ in the first place.)

Given the $\alpha_n$ eigenvalues of $PV$, given $P^2=I$,  and recalling that we have already shown that $\langle R_n|V| R_m\rangle=\delta_{m,n}$, we can thus set
\begin{eqnarray}
\langle R_n|P| R_m\rangle&=&\alpha_m^{-1}\langle R_n|PPV| R_m\rangle
\nonumber\\
&=&\alpha_m^{-1}\langle R_n|V| R_m\rangle=\alpha_m^{-1}\delta_{m,n}.
\label{N26}
\end{eqnarray}
Thus when $P$ effects $P^{-1}HP=H^{\dagger}$ we establish that $\langle R_n|P| R_m\rangle$ is time independent even though $H$ does not commute with $P$. However, even though $\langle R_n|V| R_n\rangle$ is positive definite, $\langle R_n|P| R_n\rangle$ is not required to be positive definite since the $\alpha_n$ are not in general positive definite.  On inserting a complete set of position eigenstates we obtain
\begin{eqnarray}
&&\langle R_n|P| R_n\rangle=\int dx\langle R_n|P|x\rangle\langle x| R_n\rangle
\nonumber\\
&&=\int dx\langle R_n|-x\rangle\langle x| R_n\rangle=\int dx\psi_n^*(-x)\psi_n(x).~~
\label{N27}
\end{eqnarray}
Now at this point we are not free to normalize the $\langle R_n|P| R_n\rangle=1/\alpha_n$ matrix elements to $\pm 1$ as the normalization of the states has already been fixed by $\langle R_n^{\prime}| R_n^{\prime}\rangle=\langle R_n|V| R_n\rangle=1$.  Nonetheless, we can still make a positive definite norm out of $\langle R_n|P| R_n\rangle$ by noting that 
\begin{eqnarray}
\alpha_n\langle R_n|P| R_n\rangle=1=\alpha_n\int dx\psi_n^*(-x)\psi_n(x),
\label{N28}
\end{eqnarray}
and with this relation note that since as shown above $\int dx\psi_n^*(-x)\psi_n(x)$ is real, the $\alpha_n$ are in fact real after all. All that is required for the $\langle R_n|P| R_n\rangle$ norm is some operator $PV$ that commutes with $H$. The operator does not need to be a discrete operator such as $C$ that  squares to one. Thus even if we were to define $\int dx \psi_n^*(-x)\psi_n(x)$ as the $PT$ norm, the theory would still automatically find the $\alpha_n$ for us without ever needing to introduce $C$.

\section{Status of the $C$ Operator}

Now the utility of the $\int dx\psi_n^*(-x)\psi_n(x)$ norm is that often we cannot construct $V$ in a closed form. Thus if we start with the Schr\"odinger equation for the $\psi_n(x)$, as discussed in \cite{Bender2007} we would be led to the orthogonal but not positive definite $\int dx\psi_n^*(-x)\psi_m(x)$ norm in the $PT$ case whenever $H$ obeys $P^{-1}HP=H^{\dagger}$, with the Dirac-type $\int dx\psi_n^*(x)\psi_m(x)$ norm not actually being an orthogonal norm in such cases. (We show this in an explicit example below.) However to get a positive norm we would have to introduce the $PV$ operator rather than the $C$ operator as that is what the theory leads us to even if we cannot construct either the operator $V$ or the operator $PV$ in a closed form, and even while a $C$ operator that commutes with $H$ and squares to one will always exist  ($C$ and $H$  commute for any choice of their $c_n$ and $E_n$ eigenvalues in the basis in which they can simultaneously be diagonalized.) But if we did not know about $V$ at all we would have to introduce $C$ in order to get a positive definite inner product by replacing $\int dx\psi_n^*(-x)\psi_n(x)$ by $\int dx\psi_n^*(-x)c_n\psi_n(x)$, though actually at that point we would not specifically know whether the theory actually supports this particular $C$ based norm. We would however know that since the eigenspectrum of $H$ is real and complete, one must be able to bring $H$ to a Hermitian form by an appropriate (even if not explicitly known) similarity transform, and thus the theory would necessarily support some positive definite norm.

However, in introducing $C$ into the norm we would be introducing it from the outside, whereas $PV$ would be generated by the theory itself. Moreover, even if one is prepared to introduce an operator by hand, one would never need to use an operator with eigenvalues equal to plus or minus one at all, as one would only need to use an operator with positive or negative eigenvalues $\alpha_n$. Thus even if one starts with $\int dx\psi_n^*(-x)\psi_n(x)$ as the norm, one still does not need to introduce a $C$ that obeys $C^2=I$ in order to derive a positive-definite inner product from it.

Finally, it may be the case that $P$ does not generate $PHP=H^{\dagger}$ at all, or it may even be the case that $H$ is parity invariant, with just $T$ serving as the antilinear symmetry. Then $PV$ would not commute with $H$ and we could never get to the $\alpha_n\int dx\psi_n^*(-x)\psi_n(x)$ norm in the first place. So in this case both the $PV$ and $C$ norms would be irrelevant (even though the $C$ operator would still exist), but one could still use the $V$ norm since it always exists \cite{footnote8A}.

In the development of the $C$ operator two key properties were identified \cite{Bender2007}, namely that it obeyed
\begin{eqnarray}
[C,H]=0,\qquad C^2=I. 
\label{N29}
\end{eqnarray}
Now in the case where all energy eigenvalues are real, the eigenstates of $H$ are also eigenstates of $PT$. Hence, on recalling that the eigenvalues of $PT$ are real, in the primed basis we can set 
\begin{eqnarray}
P^{\prime}T^{\prime}C^{\prime}&=&\sum \eta_n|n^{\prime}\rangle c_n \langle n^{\prime}|
\nonumber\\
&=&\sum |n^{\prime}\rangle c_n \langle n^{\prime}|\eta_n\
=C^{\prime}T^{\prime}P^{\prime},
\label{N30}
\end{eqnarray}
and thus infer that $C$ commutes with $PT$. On the other hand  when energy eigenvalues appear in complex conjugate pairs the eigenstates of $H$ and $C$ are not eigenstates of $PT$, and so we do not have $[C,PT]=0$. Thus as noted in \cite{Bender2010}, whether or not  $C$ commutes with $PT$ is thus a diagnostic for whether eigenvalues of $H$ are real or in complex pairs.

Now the $C$ operator does have some useful properties, and unless one has an alternative to $C$, one would not want to give them up, or give up its role in serving as a diagnostic for whether energies are real or in complex pairs. However, one does have such an alternative, namely $PT$ itself. First it obeys the same properties as $C$, namely it obeys
\begin{eqnarray}
[PT,H]=0,\qquad (PT)^2=I,
\label{N31}
\end{eqnarray}
and in addition it does serve as a diagnostic for whether energies are real or in complex pairs, since as we had noted above, the structure of the eigenspectrum correlates with  whether or not eigenstates of $H$ are eigenstates of $PT$. We thus see that the antilinear $PT$ operator can not only achieve everything that the linear $C$ operator is capable of  achieving in those cases where the $C$ operator might be relevant, one can use $PT$ even in those cases in which the $C$ operator is not relevant at all, since with our definition of the $PT$ conjugate as involving the $\eta_n$ phase, the PT conjugate norm is always the same as the $V$ norm, and the $V$ norm always exists. 

To understand the ubiquity of the $V$ norm it is instructive to follow \cite{Mannheim2015b}. Thus suppose we start from scratch and look for a time-independent norm for $H$. Noting that the Dirac inner product $\langle R_n(t)|R_m(t) \rangle=\langle R_n(0)|\exp(iH^{\dagger}t)\exp(-iHt)|R_m(0) \rangle$ is not equal to $\langle R_n(t=0)|R_m(t=0) \rangle$ when the Hamiltonian is not Hermitian, in the non-Hermitian case the standard Dirac inner product is not preserved in time. To rectify this we introduce some as yet undetermined operator $V$ and look at norms of the form $\langle R_n(t)|V|R_m(t)\rangle$. For them we obtain 
\begin{eqnarray}
i\frac{\partial}{\partial t} \langle R_n(t)|V|R_m(t)\rangle
=\langle R_n(t)|(VH-H^{\dagger}V)|R_m(t)\rangle.~~
\label{N32}
\end{eqnarray}
We thus see that the $V$-based inner products will be time independent if $V$ obeys none other than the relation $VH-H^{\dagger}V=0$ introduced above. Then when $V$ is invertible,  the $V$ operator that gives rise to a time independent norm is thus none other than the intertwining operator that effects $VHV^{-1}=H^{\dagger}$ \cite{footnote9A}. 

As regards the converse,  suppose we are given that the $V$ norm is time independent. We would then obtain $\langle R_n(t)|(VH-H^{\dagger}V)|R_m(t)\rangle=0$ for all states $|R_m(t)\rangle$. Then if these states are complete and $V$ is invertible we could then set $VH-H^{\dagger}V=0$ and $VHV^{-1}=H^{\dagger}$ as operator identities. The condition $VH-H^{\dagger}V=0$ is thus both necessary and sufficient for the time independence of the $V$-based inner product. But the condition $VHV^{-1}=H^{\dagger}$ is also necessary and sufficient for the existence of an antilinear symmetry, since $H$ and $H^{\dagger}$ would then have the same set of energy eigenvalues, and as we had noted above that is a necessary and sufficient condition for antilinearity. Thus as noted in \cite{Mannheim2015b}, antilinearity is both necessary and sufficient for the time independence of inner products, with the antilinearity of a Hamiltonian thus being the most general condition for which one could construct a viable quantum mechanics, being so whether energy eigenvalues are real or in complex pairs.

In addition, we also note that in the complex energy case if  $|R_m(t) \rangle$ is an eigenstate of $H$ with energy eigenvalue $E_m=E_m^R+iE_m^I$, in general we can write
\begin{eqnarray}
&&\langle R_n(t)|V|R_m(t) \rangle
\nonumber\\
&&=\langle R_n(0)|V|R_m(0) \rangle e^{-i(E_m^R+iE_m^I)t+i(E_n^R-iE_n^I)t}.
\label{N33}
\end{eqnarray}
Since $V$ has been chosen so that the $\langle R_n(t)|V|R_m(t) \rangle$ matrix elements are to be time independent,  the only allowed non-zero matrix elements are those that obey
\begin{eqnarray}
&&E_m^R=E_n^R,\qquad E_m^I=-E_n^I,
\label{N34}
\end{eqnarray}
with all other $V$-based matrix elements having to obey $\langle R_n(0)|V|R_m(0) \rangle=0$. We recognize (\ref{N34}) as being precisely none other than the requirement that eigenvalues be real or appear in complex conjugate pairs, just as required of antilinear symmetry. Inspection of (\ref{N33}) and (\ref{N34}) also shows that in the presence of complex energy eigenvalues the time independence of the $V$-based inner products is maintained because the only non-zero overlap of any given $|R_m(t) \rangle$ with a given complex energy eigenvalue is that with the appropriate $\langle R_n(t)|$ with a complex energy with the opposite sign for the imaginary part. The only non-trivial matrix elements are thus those that connect the two states in a complex pair, with the time independence being maintained by a transition between a decaying mode and a growing one. With there thus being no transitions between a state and itself, in the complex energy case there are no diagonal matrix elements, and  there is thus no need to seek a positive definite norm since the overall signs of transition matrix elements are not constrained in quantum theory. Finally, while the $C$ operator will continue to exist in the complex energy case (the Hamiltonian still being diagonalizable if the set of complex energy eigenstates is complete, with a $C$ that commutes with it being simultaneously diagonalizable too), $C$ will play no role in fixing the sign of inner products. The utility of the $C$ operator is thus restricted to the real energy case only, though even there one should use the $PV$ operator, and one can even only use the $PV$ operator provided $P$ implements $P^{-1}HP=H^{\dagger}$. However, the $V$-based inner product can be used no matter whether energies are real or in complex pairs and regardless of whether or not $P$ implements $P^{-1}HP=H^{\dagger}$ at all. And thus the $V$-based norm is uniquely and unambiguously selected as the inner product that will always be time independent for any non-Hermitian Hamiltonian with antilinear symmetry.

It is also of interest to discuss completeness relations using the  $V$-based inner product, in order to see how they differ in the real and complex energy cases, and to this end it is instructive to introduce left- and and right-handed eigenvectors of $H$. We had noted in (\ref{N5}) that $-i\partial_t\langle R_n|=\langle R_n|H^{\dagger}$. Since $H^{\dagger}=VHV^{-1}$ we can thus set $-i\partial_t\langle R_n|V=\langle R_n|VH$. Thus if we define $\langle R_n|V=\langle L_n|$, we can identify $\langle L_n|$ as a left-eigenvector of $H$, with $|R_n\rangle$ itself being a right-eigenvector of $H$. In terms of the left- and right-eigenvectors (\ref{N13}) can be rewritten as 
\begin{eqnarray}
\langle R_n|V|R_m\rangle=\langle L_n|R_m\rangle=\delta_{m,n},
\label{N35}
\end{eqnarray}
when all energies are real. From (\ref{N35}) we immediately obtain
\begin{eqnarray}
&&\sum |R_n\rangle\langle L_n|=\sum |R_n\rangle\langle R_n|V=I,
\nonumber\\
&&H=\sum |R_n\rangle E_n \langle L_n|=\sum |R_n\rangle E_n\langle R_n|V,
\label{N36}
\end{eqnarray}
with any operator $O$ (including $C$) of the form
\begin{eqnarray}
O=\sum |R_n\rangle \alpha_n \langle L_n|=\sum |R_n\rangle \alpha_n\langle R_n|V
\label{N37}
\end{eqnarray}
with c-number $\alpha_n$ immediately commuting with $H$.

In the complex energy case we have
\begin{eqnarray}
PT|R_{\pm} \rangle=|R_{\mp} \rangle,\quad
\langle L_{\pm}|TP=\langle L_{\mp}|,
\label{N38}
\end{eqnarray}
with time dependences
\begin{eqnarray}
|R_{\pm} \rangle\sim \exp(-iE_{\pm}t)=\exp(-iE_Rt\pm E_It), 
\nonumber\\
\langle L_{\pm}|=\langle R_{\pm}|V \sim \exp(iE_{\mp}t)=\exp(iE_Rt \pm E_It).
\label{N39}
\end{eqnarray}
Thus we can set \cite{Mannheim2015b} 
\begin{eqnarray}
\langle L^{-}_{n}|R^{+}_{m}\rangle=\langle L^{+}_{n}|R^{-}_{m}\rangle=\delta_{n,m},
\nonumber\\
\langle L^{-}_{n}|R^{-}_{m}\rangle=\langle L^{+}_{n}|R^{+}_{m}\rangle=0,
\nonumber\\
\sum_n\bigg{[}|R^{+}_{n}\rangle\langle L^{-}_{n}|+|R^{-}_{n}\rangle\langle L^{+}_{n}|\bigg{]}=I,
\nonumber\\
H=\sum_n\bigg{[}|R^{+}_{n}\rangle E^{+}_{n}\langle L^{-}_{n}|+|R^{-}_{n}\rangle E^{-}_{n}\langle L^{+}_{n}|\bigg{]}.
\label{N40}
\end{eqnarray}
As we see, any operator $O$ (including $C$) of the form
\begin{eqnarray}
O=\sum_n\bigg{[}|R^{+}_{n}\rangle \alpha^{+}_{n}\langle L^{-}_{n}|+|R^{-}_{n}\rangle \alpha^{-}_{n}\langle L^{+}_{n}|\bigg{]}.
\label{N41}
\end{eqnarray}
with c-number $\alpha^+_n$ and $\alpha^-_n$ immediately commutes with $H$. Thus when energies are in complex pairs one can still construct a $C$ operator. However, its only non-vanishing elements would involve transition matrix elements, and since their overall signs  are not constrained in quantum theory, $C$ would play no role.  Moreover, even if $P$ does effect $P^{-1}HP=H^{\dagger}$, $PV$ would also play no role. It is only the $V$-based inner products that would be of significance.

\section{The Two-Dimensional Puzzle}

Now while we have shown that the $C$ operator is not always relevant even when all energies are real, in a study of matrices \cite{Bender2010} it was shown that it apparently always is. We thus need to reconcile these two results. On noting that complex energy eigenvalues always have to come in pairs in a $PT$-symmetric theory,  to explore the general structure of 
$PT$-symmetric theories first two-dimensional matrices were studied. And then it was noted that since for any diagonalizable matrix of any dimension one can always bring it to a form in which the matrix block diagonalizes into two-dimensional blocks, the results of \cite{Bender2010} thus generalized to arbitrary dimension. Moreover, since one can diagonalize a Hamiltonian in a Fock space basis, the results could even generalize to infinite dimension, with the Harmonic oscillator Hamiltonian $H=(a^{\dagger}a+1/2)\hbar \omega$  for instance being a well-defined operator in an infinite-dimensional Fock space.

Now in these two-dimensional studies we did not explicitly show that one can always have $P^{-1}HP=H^{\dagger}$ and $(PV)^2=I$. However, as we now show, it  turns out that one can. In this two-dimensional study we defined parity and time reversal as being associated with operators $P$ and $T$ that obeyed the standard $P^2=I$, $P=P^{\dagger}=P^{-1}$, $T^2=I$, $T=KU$, $UU^{\dagger}=I$ and $[P,T]=0$. In the two space this in general led to $P=\boldsymbol{\sigma}\cdot \textbf{p}$, $T=K\sigma_2
\boldsymbol{\sigma}\cdot \textbf{t}$, where $\textbf{p}=\textbf{p}^*$, $\textbf{p}\cdot\textbf{p}=1$, $\textbf{t}=\textbf{t}^*$, $\textbf{t}\cdot\textbf{t}=1$, and $\textbf{p}\cdot\textbf{t}=0$. With this structure we found that the general $H=\sigma_0h_0+\boldsymbol{\sigma}\cdot \textbf{h}$ would be $PT$ symmetric if
\begin{eqnarray}
h^I_0=0,\qquad (\textbf{h}_I\cdot\textbf{p})\textbf{p}+(\textbf{h}_I\cdot\textbf{t})\textbf{t}=0,
\nonumber\\ 
(\textbf{h}_R\cdot\textbf{p})\textbf{p}+(\textbf{h}_R\cdot\textbf{t})\textbf{t}-\textbf{h}_R=0.
\label{N42}
\end{eqnarray}
Satisfying these conditions leads to
\begin{eqnarray}
\textbf{h}_I\cdot\textbf{p}=0,\qquad \textbf{h}_I\cdot\textbf{t}=0,\qquad
\textbf{h}_R\cdot\textbf{h}_I=0.
\label{N43}
\end{eqnarray}
For our purposes here we can ignore $h^R_0$, and by rotational invariance can set $\textbf{h}_R=(\alpha,0,0)$ and  $\textbf{h}_I=(0,i\beta,0)$ where $\alpha$ and $\beta$ are both real and positive. And with these conditions we precisely obtain none other than the example given in (\ref{N4}), viz
\begin{eqnarray}
H=\alpha\sigma_1+i\beta \sigma_2=\sigma_1(\alpha\sigma_0-\beta \sigma_3).
\label{N44}
\end{eqnarray}
Without loss of generality we can take $\textbf{p}$ to be parallel to $\textbf{h}_R$ in (\ref{N42}), and thus set  $\textbf{p}=(1,0,0)$, $\textbf{t}=(0,0,1)$. We thus obtain $P=\sigma_1$, $T=Ki\sigma_1$ and $PT=Ki$, so that with $\alpha\sigma_1+i\beta \sigma_2$ being real, $H$ is indeed $PT$ symmetric.

For this $H$ we introduce
\begin{eqnarray}
S=\cosh\theta-\sigma_3\sinh\theta,~~~S^{-1}=\cosh\theta+\sigma_3\sinh\theta,
\label{N45}
\end{eqnarray}
and find that
\begin{eqnarray}
SHS^{-1}&=&(\alpha\cosh(2\theta)-\beta\sinh(2\theta))\sigma_1
\nonumber\\
&+&(\beta\cosh(2\theta)-\alpha\sinh(2\theta)i\sigma_2.
\label{N46}
\end{eqnarray}
$H$ will have real eigenvalues if $\alpha>\beta$, and $SHS^{-1}$ will then be Hermitian if
\begin{eqnarray}
\alpha\sinh(2\theta)-\beta\cosh(2\theta)=0,
\label{N47}
\end{eqnarray}
i.e. if
\begin{eqnarray}
\cosh(2\theta)=\frac{\alpha}{(\alpha^2-\beta^2)^{1/2}},~\sinh(2\theta)=\frac{\beta}{(\alpha^2-\beta^2)^{1/2}}.~~
\label{N48}
\end{eqnarray}
Under these conditions $SHS^{-1}$ is then given by 
\begin{eqnarray}
SHS^{-1}=(\alpha^2-\beta^2)^{1/2}\sigma_1,
\label{N49}
\end{eqnarray}
just as needed for a Hermitian $SHS^{-1}$ with eigenvalues $\pm (\alpha^2-\beta^2)^{1/2}$. 

Given $S$, one can show that with $V=S^{\dagger}S$ one obtains
\begin{eqnarray}
V&=&\cosh(2\theta)-\sigma_3\sinh(2\theta),
\nonumber\\
V^{-1}&=&\cosh(2\theta)+\sigma_3\sinh(2\theta),
\nonumber\\
VHV^{-1}&=&\alpha\sigma_1-i\beta \sigma_2=H^{\dagger},
\label{N50}
\end{eqnarray}
just as required. When $\alpha>\beta$ the eigenvectors of $H$ are
\begin{eqnarray}
u_+&=&\frac{1}{N_+^{1/2}}\begin{pmatrix}(\alpha+\beta)^{1/2}\\ (\alpha-\beta)^{1/2}\\ \end{pmatrix},
\nonumber\\
u_-&=&\frac{1}{N_-^{1/2}}\begin{pmatrix}(\alpha+\beta)^{1/2}\\ -(\alpha-\beta)^{1/2}\\ \end{pmatrix},
\label{N51}
\end{eqnarray}
and with $N_+=N_-=2(\alpha^2-\beta^2)^{1/2}$, are normalized as 
\begin{eqnarray}
u_+^{\dagger}Vu_+&=&1,\qquad u_-^{\dagger}Vu_-=1,
\nonumber\\
u_+^{\dagger}Vu_-&=&0,\qquad u_-^{\dagger}Vu_+=0,
\label{N52}
\end{eqnarray}
just as required of the $V$-based norm.

On evaluating the Dirac-type norm we obtain 
\begin{eqnarray}
u_-^{\dagger}u_+&=&\frac{1}{(N_+N_-)^{1/2}}\begin{pmatrix}(\alpha+\beta)^{1/2}& -(\alpha-\beta)^{1/2}\\ \end{pmatrix},
\nonumber\\
&&\times \begin{pmatrix}(\alpha+\beta)^{1/2}\\ (\alpha-\beta)^{1/2}\\ \end{pmatrix}=\frac{\beta}{(\alpha^2-\beta^2)^{1/2}} \neq 0,
\label{N53}
\end{eqnarray}
and confirm that the states are not Dirac orthogonal. Now with $P=\sigma_1$, we find that for the two-dimensional model $P$ does effect $P^{-1}HP=H^{\dagger}$. If we were to define a PT conjugate of the form $u_{\pm}^{PT}=u_{\pm}^{\dagger}P$ (i.e. without the intrinsic $PT$ phase), we would obtain a $P$ norm
\begin{eqnarray}
u_+^{\dagger}\sigma_1u_+&=&1,\qquad u_-^{\dagger}\sigma_1u_-=-1,
\nonumber\\
u_+^{\dagger}\sigma_1u_-&=&0,\qquad u_-^{\dagger}\sigma_1u_+=0,
\label{N54}
\end{eqnarray}
that is not positive definite. Noting however that $P$ effects $P^{-1}HP=H^{\dagger}$, the quantity $PV$ commutes with $H$ and obeys the following relations
\begin{eqnarray}
[PV,H]=0,\qquad PVP=V^{-1},\qquad (PV)^2=I.
\label{N55}
\end{eqnarray}
Thus now we can set $C=PV$ where $C^2=I$, and with $V=PC$, we see that the $PC$ norm is positive definite, just as required. This norm is equivalent to defining a PT conjugate of the form $u_{\pm}^{PT}=u_{\pm}^{\dagger}P\eta_{\pm}$ (i.e. with the intrinsic $PT$ phase), with the eigenvalues of $C$ acting the same way as $\eta_{\pm}$. To conclude, we see that all of the general ideas regarding norms hold in this simple model, and this then raises the question of why $P^{-1}HP=H^{\dagger}$, $(PV)^2=I$, $PV=C$ would then not always hold in any case in which  $PT$ symmetry is realized via a real and complete energy eigenspectrum.

\section{Solution to the Puzzle}

To see why the relations $P^{-1}HP=H^{\dagger}$, $(PV)^2=I$, $PV=C$ do not hold in general, we consider an infinite-dimensional space and introduce a Fock space vector $|\psi\rangle=\sum c_n|n\rangle$ as expanded in a complete set of $n$-particle Fock space states. We look for it to be an eigenstate of the position operator according to $\hat{x}|\psi\rangle=(a+a^{\dagger})|\psi\rangle=x|\psi\rangle$. And with $a|n\rangle=n^{1/2}|n-1\rangle$, $a^{\dagger}|n\rangle=(n+1)^{1/2}|n+1\rangle$, find the recurrence relation 
\begin{eqnarray}
(n-1)^{1/2}c_{n-2}+n^{1/2}c_{n}=xc_{n-1}.
\label{N56}
\end{eqnarray}
Thus we obtain
\begin{eqnarray}
c_1&=&xc_0,~c_2=c_0\frac{(x^2-1)}{2^{1/2}},~c_3=c_0\frac{(x^3-3x)}{6^{1/2}},
\nonumber\\
c_4&=&c_0\frac{(x^4-6x^2+3)}{(24)^{1/2}},
\nonumber\\
c_5&=&c_0\frac{(x^5-10x^3+15x)}{(120)^{1/2}},
\nonumber\\
c_6&=&c_0\frac{(x^6-15x^4+45x^2-15)}{(720)^{1/2}},....,
\label{N57}
\end{eqnarray}
so that
\begin{eqnarray}
\langle \psi|\psi\rangle=c_0^2\bigg{(}1+x^2&+&\frac{(x^2-1)^2}{2}+\frac{(x^3-3x)^2}{6}
\nonumber\\
&+&\frac{(x^4-6x^2+3)^2}{24}+...\bigg{)}.
\label{N58}
\end{eqnarray}
When $x=0$ we additionally have
\begin{eqnarray}
c_n&=&-c_{n-2}\frac{(n-1)^{1/2}}{n^{1/2}}=c_{n-4}\frac{(n-1)^{1/2}(n-3)^{1/2}}{n^{1/2}(n-2)^{1/2}}
\nonumber\\
&=&-c_{n-6}\frac{(n-1)^{1/2}(n-3)^{1/2}(n-5)^{1/2}}{n^{1/2}(n-2)^{1/2}(n-4)^{1/2}}=...
\label{N59}
\end{eqnarray}
so that $c_n^2$ grows faster than $1/n$. $\langle \psi|\psi\rangle$ thus diverges (overwhelmingly so for large $x$, while diverging at $x=0$ since $\langle \psi|\psi\rangle(x=0) =1+1/2+3/8+5/16+35/128+...$ diverges faster than the divergent $1+1/2+1/3+1/4+1/5...$). 

In consequence, the eigenstates of the position operator are not normalizable, just as is to be expected since position eigenstates obey $\langle x|x^{\prime}\rangle=\delta(x-x^{\prime})$ in the coordinate basis. Thus our matrix analysis fails in the infinite-dimensional case for states that are not normalizable (even as it would apply to a Hamiltonian such as $H=(a^{\dagger}a+1/2)\hbar \omega$ since its eigenstates are normalizable). Unfortunately, the non-normalizable states include the eigenstates of the position operator (and likewise the momentum operator), viz. precisely those operators on which we would like to implement space reflection. Thus in our setting $P^2=I$, $P=P^{\dagger}=P^{-1}$ in our two-dimensional example, we were giving $P$ all of the attributes of a parity operator save one, namely that it also is to implement space reflection.

It is this last attribute that provides $P$ with a spacetime connection, and forces us to infinite-dimensional spaces and non-normalizable states. And in such a situation $P$ may then not effect $P^{-1}HP=H^{\dagger}$ or $[PV,H]=0$ or $(PV)^2=I$. Nonetheless, in such cases we can still use the $V$-based inner product as the appropriate norm for a $PT$-symmetric theory since the $V$ norm always exists, even in an infinite-dimensional space. And for such cases we should take the $PT$ conjugate to be that conjugate that includes the intrinsic $PT$ phase and not the one that does not include it. And when we do include the $PT$ phase, the inner product associated with the overlap of a state with its $PT$ conjugate then coincides with the $V$-based inner product regardless of whether or not $P$ obeys $P^{-1}HP=H^{\dagger}$ or $[PV,H]=0$ or $(PV)^2=I$, and leads to an inner product that is fully acceptable.


\begin{thebibliography}{99}
\bibitem{footnote2A} While we will not be interested in Jordan-block Hamiltonians in this paper, we note that while such Hamiltonians do not themselves have a complete set of energy eigenvectors, the Hilbert spaces on which they act are themselves still complete. Consequently one can represent a Jordan-block Hamiltonian  as $H=\sum_{n,m}|n\rangle H_{nm}\langle m|$ where the $|n\rangle$ and $\langle n|$  are the basis vectors of the Hilbert space and its dual, and where the eigenvectors $|\psi\rangle$ that $H$ does have can be written as $|\psi\rangle=\sum_na_n|n \rangle$.
\bibitem{footnote6A} While it follows from the relation $i\partial_t\langle R_n|V|R_m\rangle=\langle R_n|VH-H^{\dagger}V|R_m\rangle$ that $\langle R_n|V|R_m\rangle$ is time independent if $VH-H^{\dagger}V=0$, it does not follow from this that $V$ is necessarily invertible. and that $VHV^{-1}=H^{\dagger}$. In this paper we restrict to invertible $V$ operators  that do effect $VHV^{-1}=H^{\dagger}$.
\bibitem{footnote7A} While we study the implications of $P^{-1}HP=H^{\dagger}$ here, we should note that its use is somewhat limited, since despite its appearance it does not follow that the operator $\hat{x}$ is necessarily parity odd. Specifically, if one descends from a relativistic theory $\hat{x}(t)$ would be the non-relativistic limit of some spin zero quantum field $\hat{\phi}(\textbf{x},t)$. $\hat{x}(t)$ would then have the same scalar or pseudoscalar parity as $\hat{\phi}(\textbf{x},t)$, and could thus have even or odd parity. In addition, we note that to establish the representation of $\hat{p}$ as $-i\partial/\partial x$ one only needs to appeal to the canonical commutation relation $[\hat{x},\hat{p}]=i$ and to use of the translation operator $\hat{T}=\exp(-ia\hat{p})$. Specifically, from the commutation relation we obtain $[\hat{x},\hat{T}]=a\hat{T}$, and for states that obey $\hat{x}|x \rangle=x|x\rangle$ we can thus set $\hat{x}\hat{T}|x\rangle=(x+a)\hat{T}|x\rangle$ and identify $\hat{T}|x\rangle=|x+a\rangle$. For small $a$ we thus obtain $\delta(x^{\prime}-x^{\prime \prime}-a)=\langle x^{\prime}|x^{\prime \prime}+a\rangle=\langle x^{\prime}|\hat{T}|x^{\prime \prime}\rangle=\delta(x^{\prime}-x^{\prime \prime})-ia\langle x^{\prime}|\hat{p}|x^{\prime \prime}\rangle$, and thus $\langle x^{\prime}|\hat{p}|x^{\prime \prime}\rangle=-i[\delta(x^{\prime}-x^{\prime \prime})-\delta(x^{\prime}-x^{\prime \prime}-a)]a=-id\delta(x^{\prime}-x^{\prime \prime})/dx^{\prime}=+id\delta(x^{\prime}-x^{\prime \prime})/dx^{\prime \prime}$. Thus even if $\hat{x}$ is parity even, as a differential operator $\hat{p}$ acts with opposite signs on $\langle x|\psi\rangle$ and $\langle \psi |x\rangle$, respectively as $-i\partial/\partial x$ and $+i\partial/\partial x$. Since $\hat{p}$ does act on bras this way, we must use the $+i\partial/\partial x$  differential form for $\hat{p}$ when it acts on them even if the Hamiltonian as a whole is parity even and $P$ does not effect $P^{-1}HP=H^{\dagger}$.
\bibitem{footnote9A} Since the discussion here starts from scratch and does not make any reference to the connection $V=S^{\dagger}S$ between $V$ and the $S$ that implements $SHS^{-1}=H^{\prime}=H^{\prime \dagger}$, the $V$-based inner product given in (\ref{N32}) will actually be time independent if $VH=H^{\dagger}V$, regardless of whether or not $V$ is invertible.
\end{thebibliography}
\end{document}